\documentclass[aps,pra,twocolumn,superscriptaddress,floatfix]{revtex4}
\usepackage{amsmath}
\usepackage{amsfonts}
\usepackage{amssymb}
\usepackage{graphicx}
\usepackage{cancel}
\usepackage[switch,columnwise]{lineno}
\usepackage{titlesec}
\usepackage{mathrsfs}

\makeatletter
\renewcommand{\section}{\@startsection{section}{1}{0mm}
  {-\baselineskip}{0.5\baselineskip}{\bf\leftline}}

\renewcommand{\subsection}{\@startsection{section}{1}{0mm}
  {-\baselineskip}{0.5\baselineskip}{\bf\leftline}}
\makeatother

\begin{document}

\title{Quantum correlation near the exceptional point}
\author{Wanxia Cao}%
\thanks{These authors contributed equally}%
\affiliation{Department of Physics, State Key Laboratory of Surface Physics and Key Laboratory of Micro
and Nano Photonic Structures (Ministry of Education), Fudan University, Shanghai 200433, China}%
\author{Xingda Lu}%
\thanks{These authors contributed equally}%
\affiliation{Department of Physics, State Key Laboratory of Surface Physics and Key Laboratory of Micro
and Nano Photonic Structures (Ministry of Education), Fudan University, Shanghai 200433, China}%
\author{Xin Meng}%
\affiliation{Department of Physics, State Key Laboratory of Surface Physics and Key Laboratory of Micro
and Nano Photonic Structures (Ministry of Education), Fudan University, Shanghai 200433, China}%
\author{Jian Sun}%
\affiliation{Department of Physics, State Key Laboratory of Surface Physics and Key Laboratory of Micro
and Nano Photonic Structures (Ministry of Education), Fudan University, Shanghai 200433, China}%
\author{Heng Shen}%
\email{heng.shen@physics.ox.ac.uk}
\affiliation{Clarendon Laboratory, University of Oxford, Parks Road, Oxford, OX1 3PU, UK}%
\author{Yanhong Xiao}%
\email{yxiao@fudan.edu.cn}
\affiliation{Department of Physics, State Key Laboratory of Surface Physics and Key Laboratory of Micro
and Nano Photonic Structures (Ministry of Education), Fudan University, Shanghai 200433, China}%
\affiliation{Collaborative Innovation Center of Advanced Microstructures, Nanjing 210093, China}

\begin{abstract}
Recent advances in non-Hermitian physical systems have led to numerous novel optical phenomena and applications. However, most realizations are limited to classical systems and quantum fluctuations of light is unexplored. For the first time, we report the observation of quantum correlations between light channels in an anti-symmetric optical system made of flying atoms. Two distant optical channels coupled dissipatively, display gain, phase sensitivity and quantum correlations with each other, even under linear atom-light interaction within each channel. We found that quantum correlations emerge in the phase unbroken regime and disappears after crossing the exceptional point. Our microscopic model considering quantum noise evolution produces results in good qualitative agreement with experimental observations. This work opens up a new direction of experimental quantum nonlinear optics using non-Hermitian systems, and demonstrates the viability of nonlinear coupling with linear systems by using atomic motion as feedback.
\end{abstract}
\maketitle

Open systems with dissipation have been a subject of growing interest in physics, and can be described by non-Hermitian Hamiltonians, in contrast to standard quantum mechanics requiring Hermiticity. Exceptional points (EP), known as non-Hermitian degeneracies or branch points, correspond to points in parameter space at which the eigenvalues of the underlying system and the corresponding eigenvectors simultaneously coalesce \cite{Moiseyev,Ganainy,Feng,Miri}. As a burgeoning research area, it gives rise to valuable insight into new classes of phenomena that are by nature difficult to address utilizing standard Hermitian representations. Exceptional points were subsequently observed in optical microcavities \cite{Lee,Zhu,Peng}, coupled atom-cavity systems \cite{Choi}, photonic crystal slabs \cite{Zhen}, exciton-polariton billiards \cite{Gao}, parity-time-symmetric systems \cite{Ruter,Regensburger,Peng2}, flying atoms \cite{antiPT} and acoustic systems \cite{Ding}. A number of intriguing physical effects related to exceptional points have been observed, such as loss-induced transparency \cite{Guo}, unidirectional invisibility \cite{Peng2,Lin}, band merging \cite{Zhen,Makris}, topological chirality \cite{Doppler,Xu}, laser mode selectivity \cite{Hodaei,Feng2},  and potentially enhanced sensitivity \cite{Hodaei2,Chen}.

A fascinating aspect of EP related to parity-time symmetry is that a large class of non-Hermitian Hamiltonians can exhibit entirely real spectra if they commute with the PT operator \cite{Bender}. At EP, PT symmetry spontaneously breaks down and the system undertakes a new phase admitting complex eigenvalues. However, the applications of PT symmetry to optical systems so far have largely relied on effective medium theories, where processes such as stochastic quantum jumps are neglected. Therefore, non-Hermitian effects in small-scale devices as well as in atomic and molecular systems are still a topic under intense investigation, where quantum processes are known to play a significant role. Earlier theoretical studies predict that the presence of quantum noise leads to significantly different physics as compared to that expected from semi-classical approaches, such as unconventional transitions from high-noise thermal emission to a coherent lasing state \cite{Kepesidis}. Such new features provide a striking route to engineer the transitions of quantum dynamical scenario by controlled dephasing interactions, but experiments with direct measurement of quantum noises near EP remain elusive.
\begin{figure*}
\centering
\includegraphics[width=0.9\textwidth]{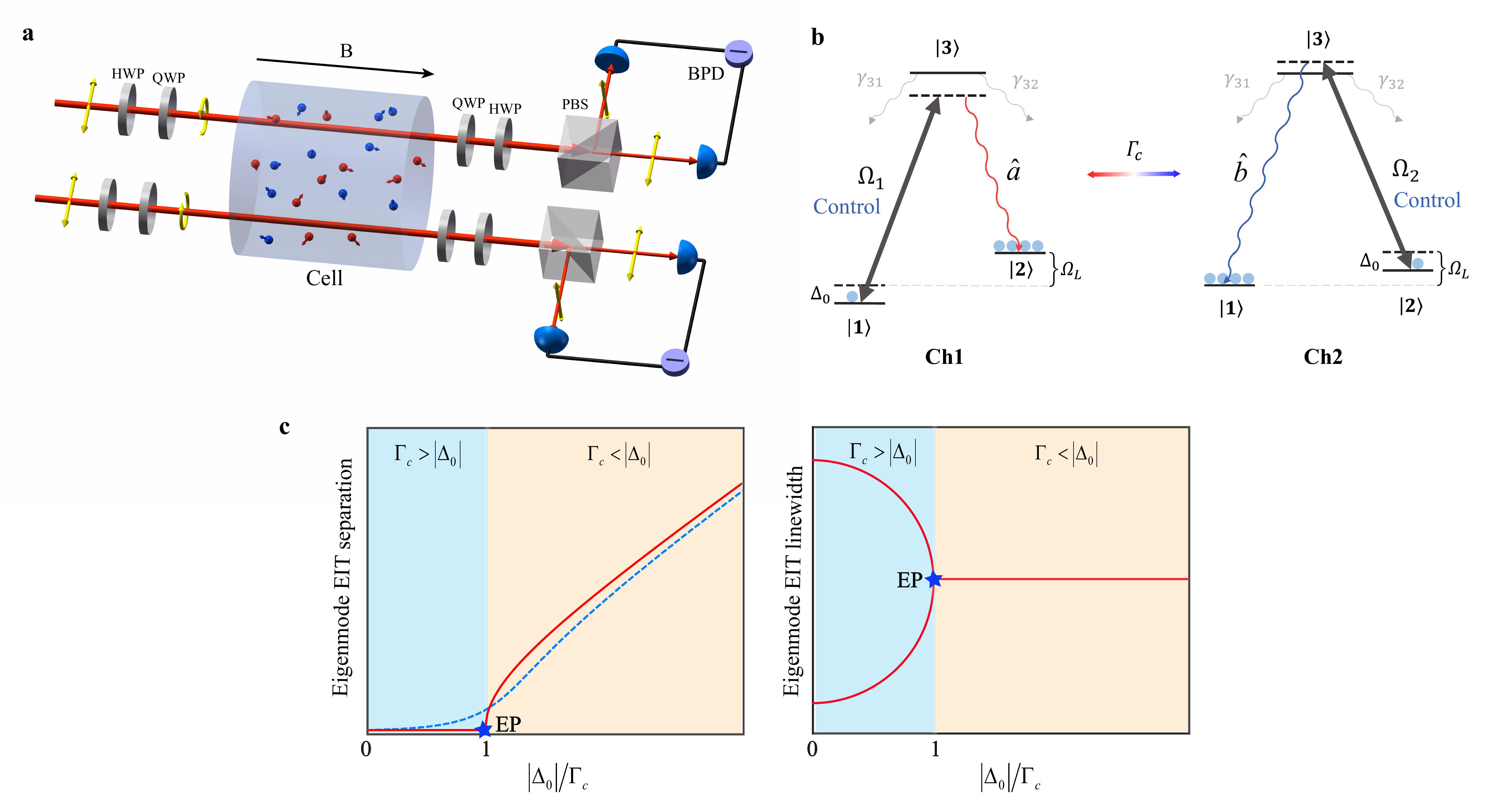}
\caption{\label{Fig:Setup} \textbf{Schematics for observation of quantum correlation in an anti-PT-symmetry platform}. \textbf{a}. the experiment schematics. Two spatially separated optical channels (Ch1 and Ch2) with the orthogonal circular polarization propagate in the warm paraffin-coated ${}^{87}$Rb vapor cell under EIT interaction. The coupling between them is mediated by coherent mixing of the atomic population and coherence of ground states created in each channel through atomic diffusion. The cell is mounted inside a four-layer magnetic shielding. Inside the shield a solenoid gives precise control over the internal longitudinal magnetic field. After the cell, the output beams are re-collimated and directed to the polarization homodyne detection setup. The noise power of the amplified subtracted photocurrents is recorded with a spectrum analyzer. PBS: polarization beam splitter; BPD: balanced photodetector; -: subtractor; HWP: half-wave plate; QWP: quarter-wave plate. \textbf{b}. The $\Lambda$ three-level scheme. The ground states are Zeeman sublevels of $\left |F=2 \right \rangle$ , and the excited state is $\left |F=1 \right \rangle$ of the ${}^{87}$Rb D1 line. The left- and right-circularly polarized control fields in Ch1 and Ch2 drive the dipole transitions $\left | 1  \right \rangle\rightarrow\left | 3  \right \rangle$ and $\left | 2  \right \rangle\rightarrow\left | 3  \right \rangle$, with Rabi frequency $\Omega_1$ and $\Omega_2$ respectively. The Zeeman splitting $\Omega_L$ is induced by a common longitudinal magnetic field, serving as either the Larmor frequency in the noise spectra measurement or the two-photon detuning in the EIT measurement (denoted as $\delta_B$ as in Fig.2). Fictional magnetic fields of opposite sign in Ch1 and Ch2 are respectively applied to shift the spin wave's frequency by $\left | \Delta_0 \right |$, via additional off-resonant laser beams (not shown). \textbf{c}. Illustration of the characteristics of the real and imaginary parts of the two EIT eigenmodes as a function of the ratio of $\left | \Delta_0 \right |$ and the coupling rate $\Gamma_c$. Exceptional point is at $\left | \Delta_0 \right |=\Gamma_c$ where the difference between the two eigen-EIT frequencies and linewidth changes abruptly. In our experiment, instead of extracting the eigenmodes, we directly observe Ch1 and Ch2's EIT frequency difference (illustrated as blue dashed line in Fig.1c) which gives a less-sharp bend at EP. }
\end{figure*}

In this regard, to investigate quantum noise near EP, we utilize a variation of the anti-PT symmetry platform of flying atoms \cite{antiPT}. The thermal motion of atoms with long-lived ground-state coherence mediates the coupling between two spatially-separated laser beams (optical channels), within which two spin waves are created by a linear process as that in a quantum memory. Quantum correlation between the two beams is built up when the two spin waves with their own distinct frequency are nearly synchronized by the atomic-motion-induced dissipative coupling. The two distant optical channels form one closed interaction loop, analogous to that of four-wave-mixing in conventional nonlinear optics. Combining the quantum light storage and coherence transport of polaritons \cite{Firstenberg}, this platform is unique for quantum studies of non-Hermitian optics.

In our system, each channel (Ch1 and Ch2) contains collinearly propagating weak-probe and strong-control fields forming electromagnetically induced transparency (EIT). The dynamics of the two collective spin-waves, i.e., the ground state coherence $\rho_{12}$ is governed by the following effective Hamiltonian \cite{antiPT},
\begin{equation}
H=\bigl(\begin{smallmatrix}
 \left | \Delta_0 \right |-i\gamma_{12} & i\Gamma_c \\
 i\Gamma_c & -\left | \Delta_0 \right |-i\gamma_{12}
\end{smallmatrix}\bigr),
\end{equation}
with its eigenvalues corresponding to the two eigen-EIT supermodes $\omega_{\pm }=-i\gamma_{12}\pm \sqrt{\Delta_0^2-\Gamma_c^2}$, where the real part corresponds to the EIT centre and the imaginary the linewidth. Here, $\left | \Delta_0 \right |$ is half the frequency difference between the two spin waves. $\gamma_{12}=\gamma_0+\Gamma_c+2\Gamma_P$ is the power broadened decay rate with $\gamma_0$ the ground-state coherence, $\gamma_c$ the ground state-coherence coupling rate between the two channels, and $2\Gamma_P=2\left | \Omega_1 \right |^2/\gamma_{13}$ the total pumping rate by the two control beams with the same Rabi frequency $\Omega_1$, where $\gamma_{13}$ is the atomic optical coherence decay rate. In such scenario (Fig.\ref{Fig:Setup}c), anti-PT symmetry breaking occurs at the exceptional point $\left | \Delta_0 \right |=\Gamma_c$ where the two supermodes perfectly overlap. In the symmetry-unbroken regime ($\left | \Delta_0 \right |<\Gamma_c$), the two eigen-EIT resonances coincide but with different linewidths. When $\left | \Delta_0 \right |>\Gamma_c$, the driven system enters the symmetry broken regime, and the resonances bifurcate and exhibit level anticrossing, resembling a passively coupled system.

\begin{figure}
\centering
\includegraphics[width=0.5\textwidth]{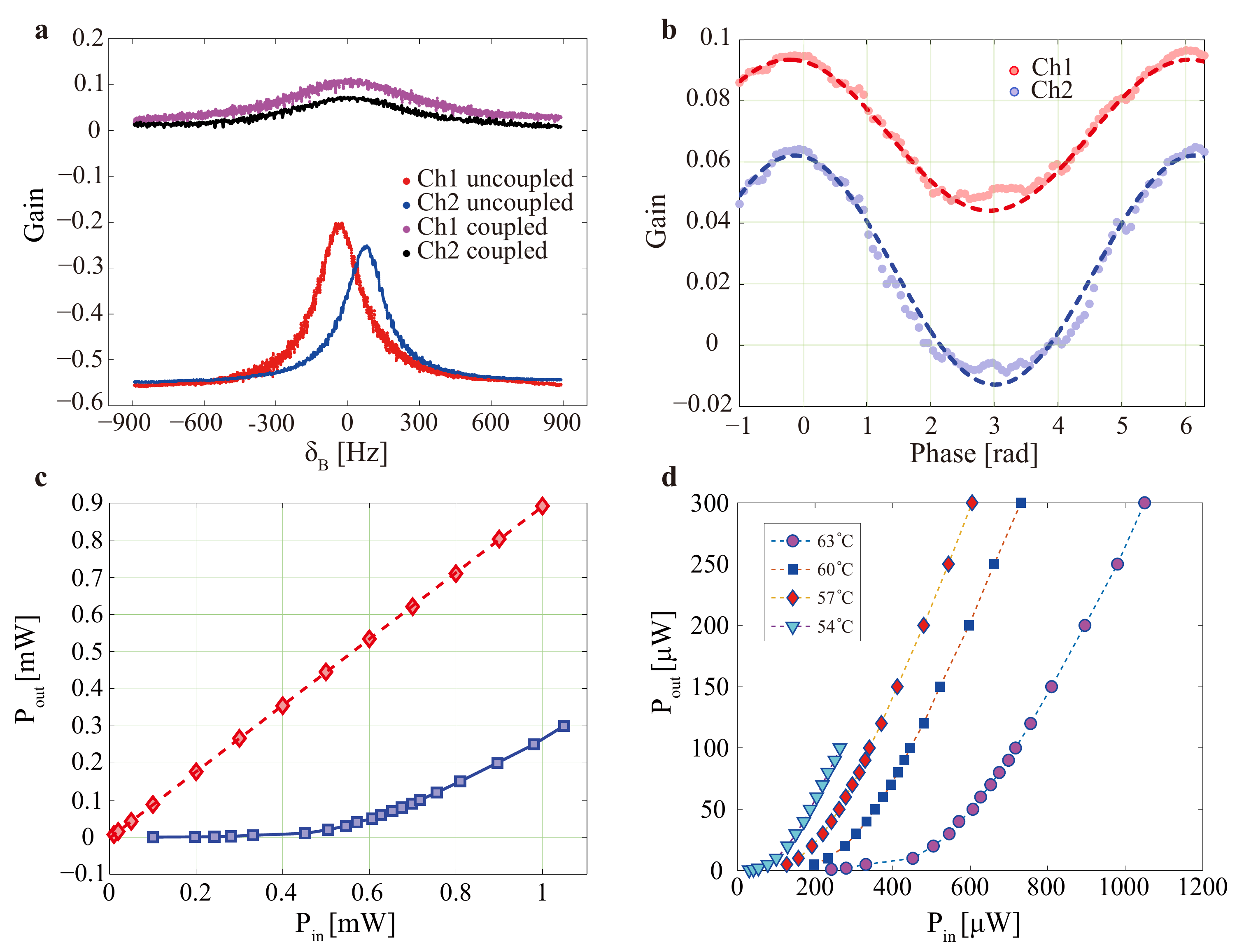}
\caption{\label{Fig:nonlinear} \textbf{Effective nonlinear interaction between two weak-probe fields aided by anti-PT-symmetric coupling of spin waves}. \textbf{a}. Probe gain as a function of two-photon detuning $\delta_B$, proportional to the common magnetic field applied, under coupled (both channels on) and uncoupled (only one channel on) conditions. Here, "Gain" is defined by the ratio of the probe’s output power and input power (calibrated at output with a far-off-resonance probe of same input power) minus one. Positive (negative) values of "Gain" stand for gain (absorption).  \textbf{b}. Gain for the probes when the probe's phase (in radian) in one channel is swept. In \textbf{a} and \textbf{b}, the cell temperature is at $40^{\circ}C$ . \textbf{c}. Optical transmission at $65^{\circ}C$  as a function of input control power (probes off). The red diamond represents the control's transmission for single channel case (the other channel off), while the blue square is that with the other channel control on. \textbf{d}. Control's transmission as a function of input power at different temperatures, with both channel's control beams on. Threshold behaviour is more prominent at higher temperature which corresponds to a higher optical depth favourable for nonlinear interaction.}
\end{figure}
Our experiment contains a warm ${}^{87}$Rb vapour cell, with paraffin wall-coating to preserve the quantum state of the atomic spins during wall collisions. It resides inside a 4-layer magnetic-shielding and is surrounded by a set of coils to generate a homogeneous bias magnetic field along the light propagation direction. As illustrated in Fig.\ref{Fig:Setup}b, each channel undergoes the $\Lambda$-type EIT but with reversed polarization configurations for the control and the probe. To fully account for the microscopic nature of our system, we theoretically describe the spin dynamics of the moving atoms by a set of coupled differential equations, taking into account the atomic state exchange between the two laser-illuminated regions and the dark region outside the laser beams, as well as the Langevin noises.

We start with the demonstration of nonlinear coupling between the two EIT probes as theoretically predicted \cite{antiPT}, although each channel by itself is linear where gain should not appear.  As the atomic coherence is being created in Ch1, the population is driven from $\left |1\right \rangle$ to $\left |2\right \rangle$; when this atom enters Ch2, the population is then brought back to $\left |1\right \rangle$. The process is continuously boosted along with coherence transfer. As a result, both probes are coherently amplified. Mathematically, we confirmed that the nonlinear coupling between the two probes here is similar to that between the signal and idler beam in a conventional four wave mixing process, except that the coupling term here is dissipative rather than coherent. Fig.\ref{Fig:nonlinear}a and Fig.\ref{Fig:nonlinear}b show the observation of probe gain and its phase sensitivity, respectively. First, the EIT spectra were measured with only one channel on. Their centres are offset from each other (Fig.\ref{Fig:nonlinear}a) due to the opposite AC stark shift induced by the orthogonally polarized control beams interacting off-resonantly with the other upper excited state (not shown in Fig.\ref{Fig:Setup}b). In contrast, with both channels on, the dissipative coupling leads to the overlapping of two EIT spectra. Remarkably, such a behaviour illustrates the synchronization \cite{Thompson} of two spin waves coupled by atomic motion. The width is broadened due to twice the optical pumping rate and the dissipative coupling. The $\sim 10\%$ gain in the probe above the $100\%$ transmission is consistent with our theoretical calculation. Second, to verify the gain phase-sensitivity, we vary the relative phase between the control and probe only in Ch1, and found that the gain in both channels varied with a period of  $2\pi$, in agreement with the theory prediction.

Another compelling evidence of establishing the nonlinear coupling is the observation of the threshold behaviour in the control fields' transmission, in absence of the input probes. As illustrated in Fig.\ref{Fig:nonlinear}c, when only one channel is on, the output power is a linear function of the input power, however, when both are on, a threshold occurs. This is due to the competition between optical pumping and the two-channel's coupling: because the helicity of the two control beams are orthogonal, the dissipative coupling counteracts the optical pumping and therefore increases the absorption. When the optical pumping dominates over the coupling, absorption decreases and the linear behaviour reappears. Similar to other nonlinear processes, at lower temperature, such threshold behaviour is much less prominent (Fig.\ref{Fig:nonlinear}d).

We begin our exploration in the quantum regime by removing the input probes and letting them be the vacuum. Joint polarization homodyne detection is operated to extract the information of the quantum fluctuations in the output probe, as well as the quantum correlations between the two channels. A common bias magnetic field is applied to shift the homodyne measurement from DC to around the Larmor frequency, bypassing low frequency technical noises. To characterize the dynamics of quantum correlation between the two channels, Gaussian discord is evaluated via the measurement of bipartite covariance matrix (CM) \cite{Adesso,Paris}. Here, canonical position and momentum operators of the quantum light fields can be defined through the light Stokes operators as $\hat{X}=\hat{S_x}/\sqrt{\left | S_z \right |}$ and $\hat{P}=\hat{S_y}/\sqrt{\left | S_z \right |}$.

\begin{figure*}
\includegraphics[width=0.9\textwidth]{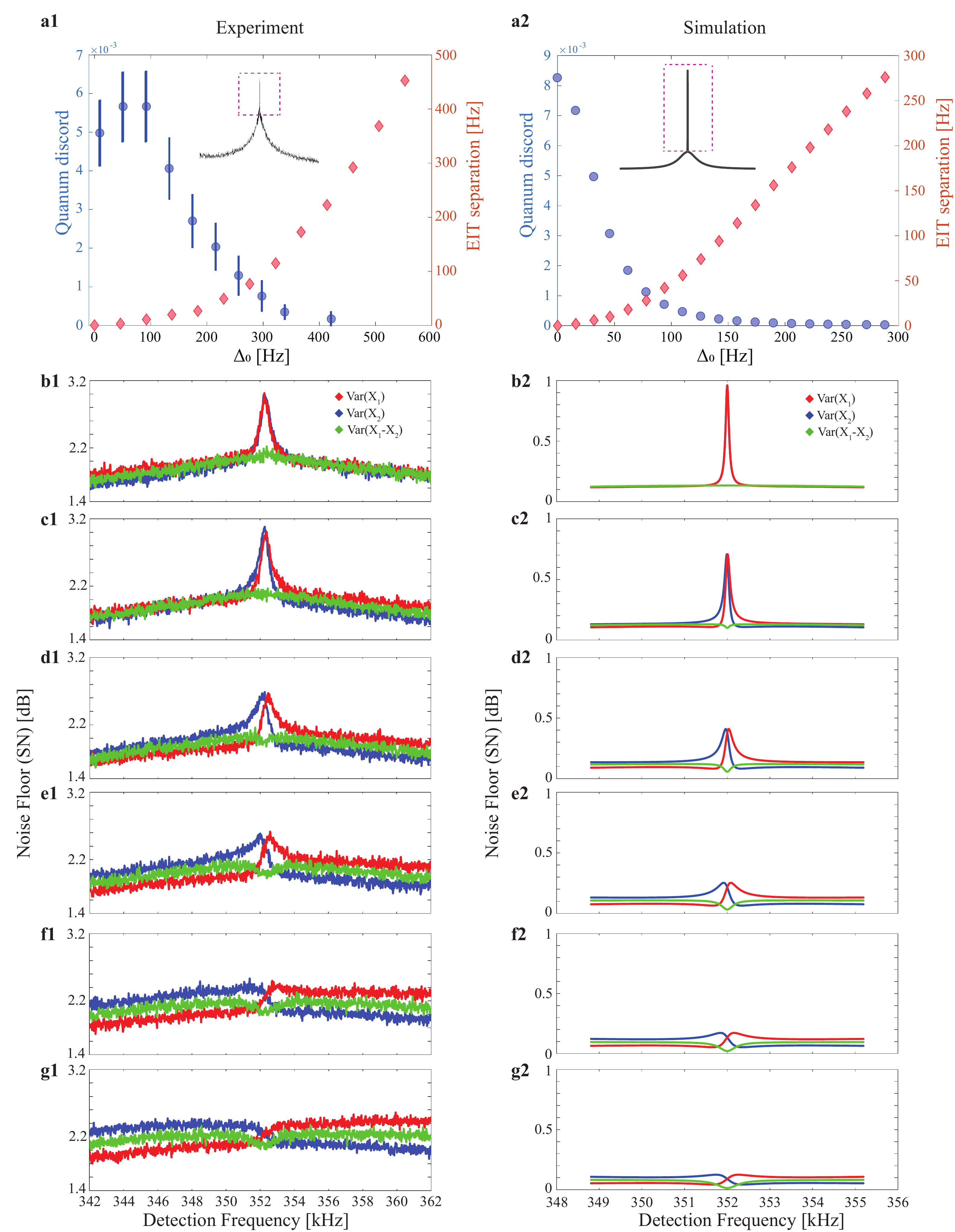}
\caption{\label{Fig:discord} \textbf{Discord, EIT separation, and noise spectra at varying $\left | \Delta_0 \right |$, the frequency offset of both spin waves but towards opposite directions as shown in Fig.1b}. The left column is experiment and the right is theory results. \textbf{a}. The discord (left-y-axis) and the separately measured EIT separation (right-y-axis) are shown at various $\left | \Delta_0 \right |$ values. The narrow structure (in dashed box) of the noise spectrum shown in the inset is displayed in b-g for sampled $\left | \Delta_0 \right |$, where the red and blue traces are $\text{Var}(\hat{X}_{1})$ and $\text{Var}(\hat{X}_{2})$ respectively, and the green is $\text{Var}(\hat{X}_1-\hat{X}_2)$. They are identical as $\text{Var}(\hat{P}_{1})$, $\text{Var}(\hat{P}_{2})$ and $\text{Var}(\hat{P}_1+\hat{P}_2)$ respectively which are not shown. The experiment noise spectra \textbf{b1}-\textbf{g1} correspond to the 1st, 3rd, 5th, 7th, 9th, 10th point on the discord curve in \textbf{a1}. The theory noise spectra \textbf{b2}-\textbf{g2} correspond to the 1st, 3rd, 5th, 7th, 9th, 11th point on the discord curve in \textbf{a2}. The shot noise level is set at 0 dB in both columns. As expected, when $\left | \Delta_0 \right |$ increases, the two spin waves are not synchronized, responsible for decreased contrast of the narrow structure. The discord drop is more dramatic near EP, where the EIT separation curve bends.  The slight (abnormal) drop in discord for the left two points in \textbf{a1} is due to the unwanted optical pumping of the off-resonant beams which changes $\left | \Delta_0 \right |$ through ac-stark shift. This optical pumping destroys some coherence and slightly decreases the contrast of the noise spectrum, as can be seen from b1 compared to \textbf{c1}.  The input power of the control in each channel is $700\mu W$ in the experiment. }
\end{figure*}

To study the change of the discord near EP, we vary the frequency difference of the two spin waves  $2\left | \Delta_0 \right |$  by applying a far-off-resonant laser beam in each channel. They create fictional magnetic fields with opposite signs via AC Stark shift because of their orthogonal polarizations. The noise spectra of the probes and the Gaussian discord at different $\left | \Delta_0 \right |$ are shown in Fig.\ref{Fig:discord}, representing the main results of this work. Fig.\ref{Fig:discord}a inset is a typical noise spectrum of each channel, which includes a broad feature originating from single pass atom-light interaction, and a sharp peak from multiple returns of atoms back to the beam. This narrow structure arises from the coupling between the two channels, and is the feature under investigation here. Measured noise spectra of each channel $\text{Var}(\hat{X}_{1,2})$, and the joint variance $\text{Var}(\hat{X}_1-\hat{X}_2)$, for six representing $\left | \Delta_0 \right |$ values are displayed in Fig.\ref{Fig:discord}b1-g1, accompanied by the theoretically calculated spectra in Fig.\ref{Fig:discord}b2-g2.  $\text{Var}(\hat{P}_{1,2})$, and $\text{Var}(\hat{P}_1+\hat{P}_2)$ are not shown since they are identical to $\text{Var}(\hat{X}_{1,2})$, and $\text{Var}(\hat{X}_1-\hat{X}_2)$ respectively. Discord is calculated at the Larmor frequency in the noise spectra. A nonzero discord value indicates the quantum nature of the correlation between the two channels.

Around EP, we observe apparent changes of the Gaussian discord with respect to $\left | \Delta_0 \right |$. When the frequency offset of the two spin waves induced by opposite fictional magnetic fields is large enough, the phases of the two spin-wave are not synchronized, reducing the efficiency of mutual coherence stimulation between the two channels. Consequently, the noise spectrum is frequency shifted from the Larmor frequency, and also becomes dispersive-like. As expected, the contrast of the narrow peak decreases, accompanied by some broadening caused by dephasing due to the differential fictional magnetic fields. These together cause the value of quantum discord to drop. When the frequency offset $\left | \Delta_0 \right |$ is smaller than $\Gamma_c$, , the system is in the phase unbroken regime, and the two spin waves' frequencies are pulled together, giving rise to relatively larger discord. To verify that the relatively sharp change in discord happens near EP, we independently measured the peak separation between the two channels' EIT with the weak probes on, and found that the separation versus $\left | \Delta_0 \right |$ curve bends around the region where discord drops fast, which corresponds to $\left | \Delta_0 \right |\sim\Gamma_c$. The corresponding theoretical curves show the same features as described above, although the linewidth of the noise spectra and $\Gamma_c$ are smaller than those in the experiments, mainly due to the neglect of multi-level effects in the model.

The results reported here, namely, observation of quantum correlations between distant optical modes under anti-PT symmetry and its relation to EP, together with the atomic-motion-enabled nonlinearity and spin-wave synchronization \cite{Thompson}, constitute a significant and important first step towards a general understanding of non-Hermitian systems with (anti-)PT-symmetry operated in the quantum regime. Extensions of our approach is possible to quantum studies in other PT or anti-PT systems currently under intense investigation, especially the microscopic ones, such as the most recently introduced solid-state spin system \cite{Murch}, as well as cold atoms \cite{Rocca}.

The authors are grateful to Klaus M\o lmer and Eugene S. Polzik for fruitful discussions. This work is supported by National Key Research Program of China under Grant No. 2016YFA0302000 and No. 2017YFA0304204, and NNSFC under Grant No. 61675047 and No. 91636107. H. Shen acknowledges the financial support from the Royal Society Newton International Fellowship (NF170876) of UK.

\end{document}